\documentclass{elsart}

\usepackage[]{graphicx}
\usepackage{amssymb}

\begin{document}

\begin{frontmatter}

\title{The spin-1 two-dimensional $J_1$-$J_2$  Heisenberg antiferromagnet on a triangular lattice}

\author{P.~Rubin$^\dagger$\thanksref{mail}},
\author{A.~Sherman$^\dagger$},
\author{M.~Schreiber$^{\dagger\dagger}$}

\address{$^\dagger$Institute of Physics, University of Tartu,
Riia 142, 51014 Tartu, Estonia}
\address{$^{\dagger\dagger}$Institut f\"{u}r Physik, Technische
Universit\"{a}t, D-09107 Chemnitz, Germany}
\thanks[mail]{Corresponding author:
E-mail: rubin@fi.tartu.ee}

\begin{abstract}
The spin-1 Heisenberg antiferromagnet on a triangular lattice with the nearest- and next-nearest-neighbor couplings, $J_1=(1-p)J$ and $J_2=pJ$, $J>0$, is studied in the entire range of the parameter $p$. Mori's projection operator technique is used as a method which retains the rotation symmetry of spin components and does not anticipate any magnetic ordering. For zero temperature four second-order phase transitions are observed. At $p\approx 0.038$ the ground state is transformed from the long-range ordered $120^\circ$ spin structure into a state with short-range ordering, which in its turn is changed to a long-range ordered state with the ordering vector  ${\bf Q^\prime}=\left(0,-\frac{2\pi}{\sqrt{3}}\right)$ at  $p\approx 0.2$. For $p\approx 0.5$ a new transition to a state with a short-range order occurs. This state has a large correlation length which continuously grows with $p$ until the establishment of a long-range order happens at $p \approx 0.65$. In the range $0.5<p<0.96$, the ordering vector is incommensurate. With growing $p$ it moves along the line ${\bf Q'-Q}_1$ to the point ${\bf Q}_1=\left(0,-\frac{4\pi}{3\sqrt{3}}\right)$ which is reached at $p\approx 0.96$. The obtained state with a long-range order can be conceived as three interpenetrating sublattices with the $120^\circ$ spin structure on each of them.

\end{abstract}


\end{frontmatter}

The active interest in the two-dimensional Heisenberg model on a triangular lattice is caused by the possibility of an unconventional spin ordering in it. The calculations carried out for the classical spins and in the spin-wave approximation \cite{Jolicoeur,Chubukov} demonstrated the rich phase diagram of the model when the nearest-neighbor coupling of spins is supplemented with more distant interactions. The interest in the model was revived by the recent synthesis of layered antiferromagnets NiGa$_2$S$_4$ \cite{Nakatsuju} and AgNiO$_2$ \cite{Coldea}. In these quasi-two dimensional crystals, magnetic properties are mainly determined by Ni$^{2+}$ ions with the spin $S=1$. These ions form a triangular lattice. NiGa$_2$S$_4$ demonstrates a spin disorder down to the temperature $T\approx 0.35$~K, incommensurate short-range spin correlations and a quadratic temperature dependence of the specific heat.

In this article we consider the $S=1$ Heisenberg model on a triangular lattice, taking into account the nearest- (NN, $J_1>0$) and the next-nearest-neighbor (NNN, $J_2>0$) couplings. We use Mori's projection operator technique \cite{Mori} which retains the rotation symmetry of spin components and does not anticipate any magnetic ordering. In this approach, the spin Green's function is represented by a continued fraction. The elements of the fraction are calculated in the recursive procedure which is similar to Lanczos' orthogonalization \cite{Sherman}. We found that the $120^\circ$ spin structure, two-sublattice and incommensurate phases with the long-range order (LRO), obtained for classical spins and in the spin-wave approximation \cite{Jolicoeur,Chubukov}, are separated by phases with a short-range order (SRO). Thus, the phase diagram of the model appears to be even richer than that obtained before -- it contains four second-order transitions.

The Hamiltonian of the  model  reads
\begin{equation}\label{hamiltonian}
H=\frac{1}{2}\sum_{\bf nm}J_{\bf nm}\left(s^z_{\bf n}s^z_{\bf
m}+s^{+1}_{\bf n}s^{-1}_{\bf m}\right),
\end{equation}
where $s^z_{\bf n}$ and $s^\sigma_{\bf n}$ are the components of the
spin-1 operators ${\bf s_n}$, {\bf n} and {\bf m} label sites of the triangular lattice, $\sigma=\pm 1$. The spin-1 operators can be written as $s^z_{\bf n}=\sum_{\sigma=\pm1} \sigma |{\bf n}, \sigma \rangle\langle{\bf n},\sigma|$ and $s^\sigma_{\bf n}=\sqrt{2} (|{\bf n}, 0\rangle\langle{\bf n},-\sigma| +  |{\bf n}, \sigma\rangle\langle{\bf n},0|)$, where $|{\bf n},  \pm 1\rangle$ and $|{\bf n},0\rangle$ are site states with different spin projections. As mentioned above, we take into account the NN and NNN interactions,  $J_{\bf nm}=J_1\sum_{\bf a}\delta_{\bf n,m+a} + J_2\sum_{\bf d}\delta_{\bf n,m+d} $ with the vectors {\bf a}  and {\bf d} connecting the NN and NNN  sites. By analogy with Ref.~\cite{AFB}, where a similar model for $S=\frac{1}{2}$ on a square lattice was considered, the frustration parameter $p$ is introduced, $J_1=(1-p)J$, $J_2=pJ$, $J=J_1+J_2$. In the following we use $J$ as the unit of energy and the lattice spacing as the unit of length.

The retarded Green's function reads
\begin{equation}\label{green}
 D({\bf k}t)=-i\theta(t)\langle[s^z_{\bf k}(t),s^z_{\bf
-k}]\rangle,
\end{equation}
where $s^z_{\bf k}=N^{-1/2}\sum_{\bf n}e^{-i{\bf kn}}s^z_{\bf n}$, $N$ is the number of sites, $s^z_{\bf k}(t)=e^{iHt}s^z_{\bf
k}e^{-iHt}$ and the angular brackets denote the statistical averaging.

We exploit Mori's projection operator technique \cite{Mori,Sherman} for calculating the Fourier transform of Kubo's relaxation function,
$$(\!( s^z_{\bf k}|s^z_{\bf -k})\!)_\omega=\int_{-\infty}^\infty
dt e^{i\omega t}(\!( s^z_{\bf k}|s^z_{\bf -k})\!)_t,\quad (\!( s^z_{\bf k}|s^z_{\bf -k})\!)_t=\theta(t)\int_t^\infty dt'\langle[s^z_{\bf k}(t'), s^z_{\bf - k}]\rangle.$$
The Fourier transform of Green's function (\ref{green}) can be obtained from this relaxation function using the relation
\begin{equation}\label{gk}
D({\bf k \omega})=\omega (\!( s^z_{\bf k}|s^z_{\bf
-k})\!)_\omega -(s^z_{\bf k},s^z_{\bf -k}),
\end{equation}
where  $(A,B)=i\int_0^\infty dt\langle[A(t),B]\rangle$. In this approach,  $(\!( s^z_{\bf k}|s^z_{\bf -k})\!)_\omega$ is represented as the continued fraction
\begin{equation}\label{cfraction}
(\!( s^z_{\bf k}|s^z_{\bf -k})\!)_\omega=\frac{\displaystyle(s^z_{\bf
k},s^z_{\bf -k})}{\displaystyle \omega-E_0-\frac{\displaystyle
V_0}{\displaystyle\omega-E_1-\frac{\displaystyle V_1}{\ddots}}},
\end{equation}
where the elements $E_n$  and $V_n$  of the fraction are determined from the recursive procedure
\begin{eqnarray}
&&[A_n,H]=E_nA_n+A_{n+1}+V_{n-1}A_{n-1},\quad E_n=([A_n,H],
A_n^\dagger)\,(A_n,A_n^\dagger)^{-1},\nonumber\\[-0.5ex]
&&\label{lanczos}\\[-0.5ex]
&&V_{n-1}=(A_n,A_n^\dagger)\,(A_{n-1},
A_{n-1}^\dagger)^{-1},
\quad V_{-1}=0, \quad A_0=s^z_{\bf k},\quad n=0,1,2,\ldots\nonumber
\end{eqnarray}
The operators $A_i$ constructed in this procedure form an orthogonal set, $(A_i,A^\dagger_j)\propto\delta_{ij}$.

Using procedure (\ref{lanczos}) we get
\begin{eqnarray*}
&&E_0=(i\dot{s}^z_{\bf k},s^z_{\bf -k})(s^z_{\bf k},s^z_{\bf
-k})^{-1}=\langle[s^z_{\bf k},s^z_{\bf -k}]\rangle(s^z_{\bf k},s^z_{\bf
-k})^{-1}=0,\quad A_1=i\dot{s}^z_{\bf k},\\
&&V_0=6 J \left[ (1-p) C_1 (\gamma_{\bf k}-1)+ p \, C_d (\gamma'_{\bf k}-1) \right](s^z_{\bf k},s^z_{\bf -k})^{-1}, \\
&&E_1=(i^2\ddot{s}^z_{\bf k},-i\dot{s}^z_{\bf -k})(i\dot{s}^z_{\bf
k},-i\dot{s}^z_{\bf -k})^{-1}=0,
\end{eqnarray*}
where $\gamma_{\bf k}=\frac{1}{3}\cos(k_x)+\frac{2}{3}
\cos(\frac{k_x}{2}) \cos(\frac{k_y \sqrt3}{2})$ and  $\gamma_{\bf k}^\prime=\frac{1}{3}\cos(\sqrt{3} k_y)+\frac{2}{3}
\cos(\frac{3 k_x}{2} ) \cos(\frac{k_y \sqrt3}{2})$  in the orthogonal
system of coordinates, $C_1 = \langle  s_{\bf n}^{+1} s_{\bf n +
a}^{-1} \rangle$ and $C_d = \langle  s_{\bf n}^{+1} s_{\bf n +
d}^{-1} \rangle$  are the spin correlations  on the NN and NNN sites, respectively. At this point we interrupt the continued fraction and calculate $(s^z_{\bf k},s^z_{\bf -k})$. In the accepted approximation $V_1 \propto (A_2 ,A_2^\dagger) = 0$. From this equation we find
\begin{equation}\label{aii}
\langle[i^2\ddot{s}^z_{\bf k},-i\dot{s}^z_{\bf -k}]\rangle=36 J^2 [ (1-p) C_1 (\gamma_{\bf k}-1) + p \, C_d
(\gamma'_{\bf k}-1)]^2 (s^z_{\bf k},s^z_{\bf -k})^{-1}.
\end{equation}
The quantity $i^2\ddot{s}^z_{\bf k}$ in the left-hand side of this
equation is a sum of terms of the type $s^z_{\bf l}s^{+1}_{\bf
n}s^{-1}_{\bf m}$. Following Refs. \cite{Kondo,Shimahara91}, we use the decoupling
$$s^z_{\bf l}s^{+1}_{\bf
n}s^{-1}_{\bf m}=\left[\alpha \langle s^{+1}_{\bf n}s^{-1}_{\bf
m}\rangle(1-\delta_{\bf nm})+\frac{4}{3}\delta_{\bf nm}\right]s^z_{\bf
l}$$
for the case $\bf l \ne m,n$. Here $\alpha$ is the vertex correction. In contrast to the case $S=\frac{1}{2}$ \cite{pla2005}, the terms with $\bf l = n$   or $\bf l = m$ do not cancel each other completely. For $S=1$ the residual terms read
$$
P_{\bf l}=\frac{1}{\sqrt{2}}\sum_{\bf m} J_{\bf lm}^2
\left(
|{\bf l},+1\rangle  \langle {\bf l},0| s_{\bf m}^- -
 |{\bf m},+1\rangle  \langle {\bf m},0| s_{\bf l}^- -
 s_{\bf l}^+ |{\bf m},0\rangle  \langle {\bf m},+1| +
 s_{\bf m}^+ |{\bf l},0\rangle  \langle {\bf l},+1|
\right).
$$
We neglect these terms in following calculations taking into account that $\left\langle P_{\bf l} \right\rangle =0$.

Using this approximation for $i^2\ddot{s}^z_{\bf k}$, from
Eq.~(\ref{aii}) we find $(s^z_{\bf k},s^z_{\bf -k})$ and from
Eqs.~(\ref{gk}) and (\ref{cfraction}) we get
\begin{equation}\label{gfd}
D({\bf k}\omega)=\frac{6 J \left[(1-p)(\gamma_{\bf k}-1)C_1
 +  p (\gamma^\prime_{\bf
 k}-1) \, C_d \right]}{\omega^2-
 \omega^2_{\bf k}},
\end{equation}
where
\begin{eqnarray}\label{omega}
 \omega^2_{\bf k} &=& 36 J^2 \alpha \Bigg\{ \Bigg.
(1-p)^2 (\gamma_{\bf k}-1)
\left[\frac{C_1}{6}+C_1\gamma_{\bf k}-C_2 -\frac{2 (1-\alpha) }{9 \alpha} \right] \nonumber \\ \nonumber
&+&   \left. p^2 (\gamma^\prime_{\bf k}-1)
\left[\frac{C_d}{6}+C_d\gamma^\prime_{\bf k}  -C_2^\prime -\frac{2 (1-\alpha) }{9 \alpha} \right]   \right.  \nonumber \\
&+& p (1-p)  \left[ \left(   1-\gamma_{\bf k}  \right) \left( C^{\prime \prime} - \gamma^\prime_{\bf k} C_1  \right)    +\left(   1-\gamma_{\bf k}^\prime  \right) \left( C^{\prime \prime} - \gamma_{\bf k} C_d\right)  \right]  \Bigg. \Bigg\},
\end{eqnarray}
$ C_2 = \frac{1}{3}\left(\frac{2}{3} + \frac{1}{2}\langle s_{\bf n}^{+1}s_{\bf n+ 2 a}^{-1}\rangle +C_d + C_1\right) $,
$C^{\prime \prime} =  \frac{1}{3}\left( \langle s_{\bf n}^{+1}s_{\bf n+  a + d}^{-1}\rangle  +
\langle s_{\bf n}^{+1}s_{\bf n+ 2 a}^{-1}\rangle + C_1\right)$ and $C^{\prime }_2 = \frac{1}{3}\left(\frac{2}{3} +\frac{1}{2} \langle s_{\bf n}^{+1}s_{\bf n+ 2 d}^{-1}\rangle + C_d+ \langle s_{\bf n}^{+1}s_{\bf n+ 3 a}^{-1}\rangle\right) $. As follows from Eq.~(\ref{gfd}), the quantity $\omega_{\bf k}$ is the frequency of the spin excitations. From Eq.~(\ref{omega}) we see that this frequency  tends to
zero when $\bf k \rightarrow 0$.

To find the parameters $\alpha$, $C_1 $, $C_2$, ${C}'_2 $, $C_d$ and $C^{\prime \prime}$  in Eqs.~(\ref{gfd})
and (\ref{omega}) we use the relations connecting the spin correlations with Green's function (\ref{gfd})
\begin{equation}
\label{eq1} \left\langle {s_{\rm {\bf n}}^{+1} s_{\rm {\bf m}}^{-1} }
\right\rangle =\frac{6J}{N}\sum\limits_{\rm
{\bf k}} {e^{i{\rm {\bf k}}\left( {{\rm {\bf n}}-{\rm {\bf m}}}
\right)}\frac{(1-p) ({\gamma }_{\rm {\bf k}}-1) C_1 + p \, ({\gamma }'_{\rm {\bf k}}-1) C_d}{\omega _{\rm {\bf k}}
}\coth \left( {\frac{\omega _{\rm {\bf k}}}{2 T}} \right).}
\end{equation}
Five equations for $C_1 $, ${C}_2 $, ${C}'_2 $, $C_d$, $C^{\prime \prime}$ which are derived from Eq.~(\ref{eq1}), and the equation $\langle s^{+1}_{\bf m} s^{-1}_{\bf m} \rangle =4/3$ , which follows from the constraint  $\langle  {\bf s^{2}_{\bf m} }\rangle =2$, form the closed set of equations for calculating these parameters.

The calculations were carried out for the entire range of the frustration parameter $0 \le p \le 1$ and for the temperatures $T/J=0$ and 0.2 in a $100\times100$ lattice with periodic boundary conditions (for the case $J_2=0$ lattices up to $1000\times1000$ sites were applied). For the solution of the mentioned set of six equations we used the Optimization toolbox of the Matlab package.

Let us first consider the case  $J_2=0$ $(p=0)$. In this situation, the above equations do not contain  ${C}'_2 $, $C_d$ and $C^{\prime \prime}$. For $T=0$ the system has LRO, which manifests itself in vanishing frequencies of spin excitations at nonzero ordering vectors. As follows from Eq.~(\ref{omega}), these ordering vectors are ${\bf Q}=\left( \frac{4 \pi }{3}, 0 \right)$ and $\left(\frac{2 \pi }{3}, -\frac{2 \pi }{\sqrt{3}} \right)$ in the Wigner-Seitz cell (one of these vectors is indicated in Fig.~1; other four vectors of the corners of the cell belong to neighboring cells). These vectors correspond to the mentioned $120^\circ$  spin structure.
In the case of the LRO the parameters $\alpha$,  $C_1$ and  $C_2$ are supplemented with a new parameter -- the condensation part $C$. The equation for its determination is the condition  $\omega_{\bf Q}=0$. Solving this set of equations we found  $C \approx 0.15$, which corresponds to the sublattice magnetization  $m=\sqrt{3C/2}\approx0.47$. For this case the dispersion of spin excitations is shown in Fig.~2~(a). Thus the ground state of the $S=1$  model with NN interactions is characterized by the LRO. In accordance with the Mermin-Wagner theorem \cite{MW} for nonzero temperatures the LRO changes into SRO which manifests itself in finite values of  $\omega_{\bf Q}$. Notice that for the NN $S=\frac{1}{2}$ Heisenberg model on a triangular lattice the same approach gives the ground state with the SRO -- even for $T=0$  the frequencies  $\omega_{\bf Q}$ remain finite \cite{pla2005}.

The evolution of the zero-temperature spin-excitation spectrum with $p$  is shown in Fig.~2. In the range $0 \le p \lesssim 0.038$  the frequency of spin excitations vanishes at wave vectors $\bf Q$  and the dispersion is close to that shown in Fig.~2~(a) (see also Fig.~3). Thus in this range the system retains the $120^\circ$ N\'{e}el LRO. Notice that in the analogous classical Heisenberg model and in the spin-wave approximation based on this classical solution this phase exists in the range $0\leq p\lesssim 0.11$ \cite{Jolicoeur,Chubukov}. In our approach in the range $0.038 \lesssim p \lesssim 0.2$ the frequency of spin excitations becomes finite everywhere except for $\bf k = 0$  which points to the destruction of the LRO and the establishment of a SRO. A typical dispersion in this range is shown in Fig.~2~(b).

At $p \approx 0.2$ new zeros appear in the spectrum at the wave vectors ${\bf Q^\prime}=\left(0,-\frac{2 \pi}{\sqrt{3}}\right)$ and $\left(\pi,\pm \frac{\pi}{\sqrt{3}}\right)$ of the Wigner-Seitz cell [see Fig.~1, 2~(c) and 3]. The appearance of the new ordering vectors indicates the establishment of a new LRO phase. The spin-excitation frequencies at these wave vectors remain vanishing in the range $0.2 \lesssim p \lesssim 0.5$ that determines the domain of this phase.
This phase can be confronted with the two-sublattice metamagnetic phase of the analogous classical Heisenberg model, which in our notations exists in the range  $0.11 \leq p \leq0.5$ and has the same ordering vectors \cite{Jolicoeur,Chubukov}. This latter phase can be visualized as the ferromagnetic ordering of spins along one of the principal directions of the triangular lattice and the antiferromagnetic ordering along two others. The two complementary degenerate states are obtained from the former by interchanging the directions of the ferromagnetic and antiferromagnetic ordering. In our approach quantities averaged over statistical realizations are considered, in this instance over the mentioned three degenerate states. Therefore the spin correlation $\left\langle s^{+1}_{\bf n}s^{-1}_{\bf m}\right\rangle$ does not depend on the direction of the vector ${\bf n-m}$. With this remark taken into account one can see in Fig.~4 that in the range $0.2 \lesssim p \lesssim 0.5$ the signs of our calculated spin correlations are in compliance with the discussed ground-state spin configurations of the classical model.

When the frustration parameter $p$ exceeds 0.5 a gap appears at the vector $\bf Q'$ which points to the destruction of the LRO and the establishment of a SRO. As seen in Fig.~5, simultaneously new minima of the spin-excitation dispersion begin to shape at incommensurate wave vectors on the line connecting momenta $\bf Q'$ and their equivalents with the points ${\bf Q}_1=\left(0,\pm \frac{4 \pi}{3 \sqrt{3}}\right)$, $\left(\pm \frac{2 \pi}{3}, \pm \frac{2 \pi}{3 \sqrt{3}}\right)$ (see Fig.~1). The frequency of these new minima is small and decreases with increasing $p$. The smallness of the frequency indicates that the correlation length $\xi$ of the considered SRO is large. From Eq.~(\ref{eq1}) one can see that for small temperatures and large distances $|\bf n-m|$ the main contribution to the sum is made by a small vicinity of the dispersion minima. Using this observation one can obtain an exponential dependence of the correlations on the distance and estimate $\xi$ \cite{Sherman03}. For $p=0.52$ we found $\xi\approx 30a$ and for $p=0.55$ we determined $\xi\approx 50a$, where $a$ is the lattice spacing. It should be noted that using a large but finite lattice we introduce an upper limit for the correlation length and a lower bound for the gap at the ordering vector. For the considered large lattice this gap is extremely small in the ordered states and its magnitude is comparable with the accuracy of the used optimization procedure. A complication arises when an ordering vector is incommensurate, as it happens for $p>0.5$. In the general case this vector does not coincide with any allowed wave vector for the considered finite lattice. In this instance we are unable to determine the exact location of the ordering vector and the size of the gap, as in Fig.~5 for $p\gtrsim 0.65$. However, the tendency for the decrease of the gap is well seen in this figure and there is good reason to believe that the SRO gives way to LRO at $p\approx 0.65$. The typical dispersion of spin excitations for this range is shown in Fig.~2~(d). In the classical version of the model this incommensurate LRO phase exists in the entire range $0.5\leq p<1$ \cite{Jolicoeur,Chubukov}.

At $p \approx 0.96$ the ordering vectors reach the points ${\bf Q}_1$ [see Fig. 1, 2 (e) and 3]. The values of the frustration parameter $p \gtrsim 0.96$  correspond to the case when the interaction between NNN spins is much stronger than between nearest neighbors,  $J_2 \gg J_1$. In this situations the lattice is divided into three sublattices with the elementary translation vectors  $\bf d$. Within any of the sublattices the interaction between spins is characterized by the large parameter $J_2$, while the inter-sublattice interaction is given by the smaller parameter  $J_1$. As a result the LRO established for $p\gtrsim 0.96$  can be conceived as three interpenetrating $120^\circ$ spin structures on the sublattices. In the limit $p\rightarrow 1$ spin orientations on the different sublattices are independent of one another. This conclusion can be also made from Fig.~4 -- in this limit $C_1$, the correlation function between spins in different sublattices, vanishes. At the same time $C_d$, the correlation function between NN spins within the sublattice, tends to the value of $C_1$ at $p=0$. At the classical level, the LRO with the ordering vectors ${\bf Q}_1$ is only approached asymptotically when $p\rightarrow 1$. For any $p<1$ the incommensurate spiral is lower in energy than this former state \cite{Jolicoeur}.

As seen above, our results for the quantum Heisenberg model reproduce the three LRO phases of the analogous classical model and the spin-wave approximation based on this classical solution. We found also that in the quantum model these LRO phases are separated by phases with SRO. In these transitions, opening and closing the gaps at the ordering wave vectors occur with continuity, which is inherent in second-order phase transitions.

The above results demonstrate that the considered Heisenberg model has an incommensurate SRO phase with a large correlation length and the dispersion near the minima, which can explain the quadratic temperature dependence of the specific heat observed in NiGa$_2$S$_4$ \cite{Nakatsuju}. However, the respective ordering vectors do not coincide with those found in the crystal. Apparently for proper description of the incommensurate phase in NiGa$_2$S$_4$ one has to take into account some additional interactions such as a coupling between third neighbors \cite{Mazin,boson} or biquadratic interactions \cite{Oitmaa}.

In summary, Mori's projection operator technique was used for
investigating the excitation spectrum and spin correlations of the two-dimensional $S=1$ Heisenberg antiferromagnet on a triangular lattice taking into account the nearest- ($J_1$) and next-nearest-neighbor ($J_2$) interactions. At zero temperature the competition of these interactions leads to the appearance of four second-order phase transitions between phases with long- and short-range magnetic order. At the frustration parameter $p \equiv J_2/(J_1+J_2) \approx 0.038$  the ground state of the model is transformed from the long-range ordered $120^\circ$ spin structure with the ordering vectors ${\bf Q}=\left( \frac{4 \pi }{3}, 0 \right)$ and $\left(\frac{2 \pi }{3}, -\frac{2 \pi }{\sqrt{3}} \right)$ into a state with short-range ordering. At $p \approx 0.2$ this latter state is changed to the long-range ordered state with the ordering vectors ${\bf Q^\prime}=\left(0,-\frac{2 \pi}{\sqrt{3}}\right)$ and $\left(\pi,\pm \frac{\pi}{\sqrt{3}}\right)$. This state can be related to the two-sublattice metamagnetic state of the classical model. A further transition to a state with short-range order occurs at $p\approx 0.5$. This state has large correlation length and at $p\approx 0.65$ turns into another long-range ordered state. In the range $0.5\lesssim p \lesssim 0.96$ the ordering vectors are incommensurate. With growing $p$ they move from the vectors ${\bf Q'}$ and their equivalents to the vectors ${\bf Q}_1=\left(0,\pm \frac{4 \pi}{3 \sqrt{3}}\right)$ and $\left(\pm \frac{2 \pi}{3}, \pm \frac{2 \pi}{3 \sqrt{3}}\right)$, which are reached at $p\approx 0.96$. The resulting state can be conceived as three interpenetrating sublattices with a $120^\circ$ spin structure on each of them. With $p\rightarrow 1$ the spin correlations between the sublattices are weakened. The phases with the long-range order of the quantum Heisenberg model are similar to those of the analogous classical model. Additionally the quantum model has two phases with the short-range order which separate the long-range ordered phases.

\section*{Acknowledgements}
Authors are thankful to A.~V.~Mikheyenkov for useful comments. This work was partly supported by the ESF Grant No 6918 and by the DAAD.

\newpage
\begin{center}Figures\end{center}






\begin{figure}[bhtp]
\begin{center}
\includegraphics[height=0.8\linewidth, angle=0]
{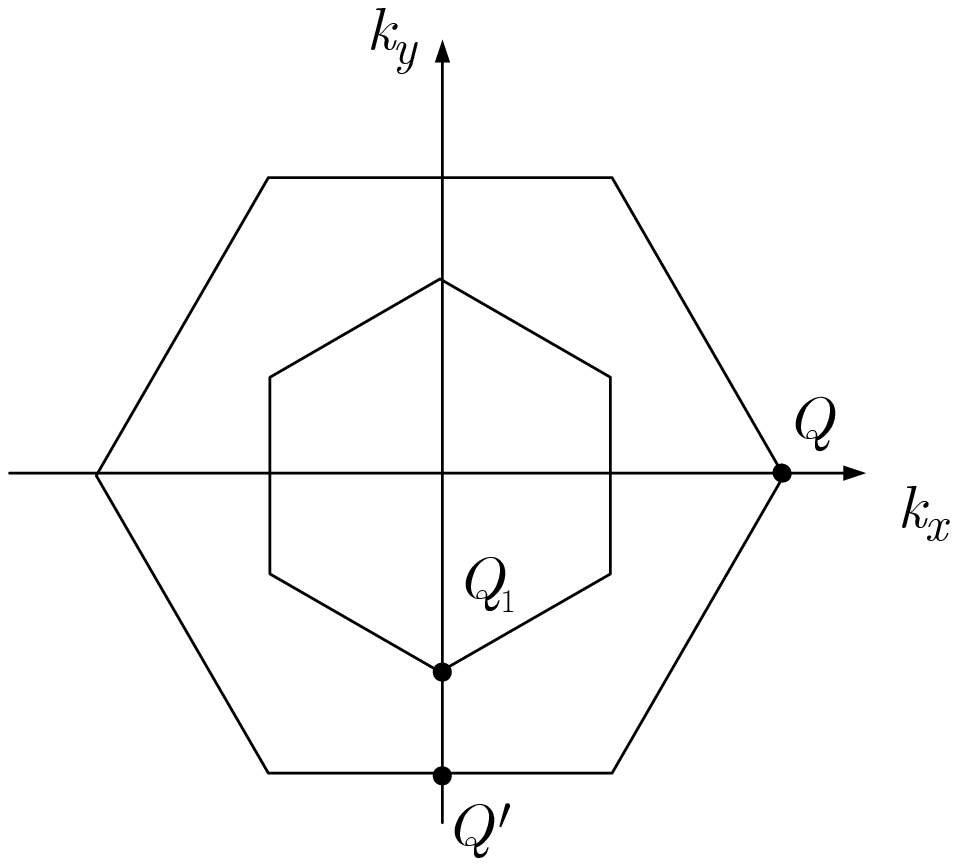}
\caption{ The Brillouin zone for the triangular lattice. ${\bf Q}=\left(\frac{4 \pi}{3},0\right)$, ${\bf Q}^\prime=\left(0, -\frac{2\pi}{\sqrt{3}}\right)$ and ${\bf Q}_1=\left( 0,-\frac{4 \pi}{3 \sqrt{3}}\right)$ are the ordering vectors of the three different LRO states of the model in the range $0 \leq p \leq 1 $. For every LRO state only one of several ordering vectors in the zone is shown. The small hexagon corresponds to the Brillouin zone of one of three interpenetrating spin sublattices with the $120^\circ$ spin structure which appear in the case $p\gtrsim 0.96$.}
\label{fig1}
\end{center}
\end{figure}
\newpage
\begin{figure}[bhtp]
\begin{center}
\includegraphics[height=1.2\linewidth, angle=0]
{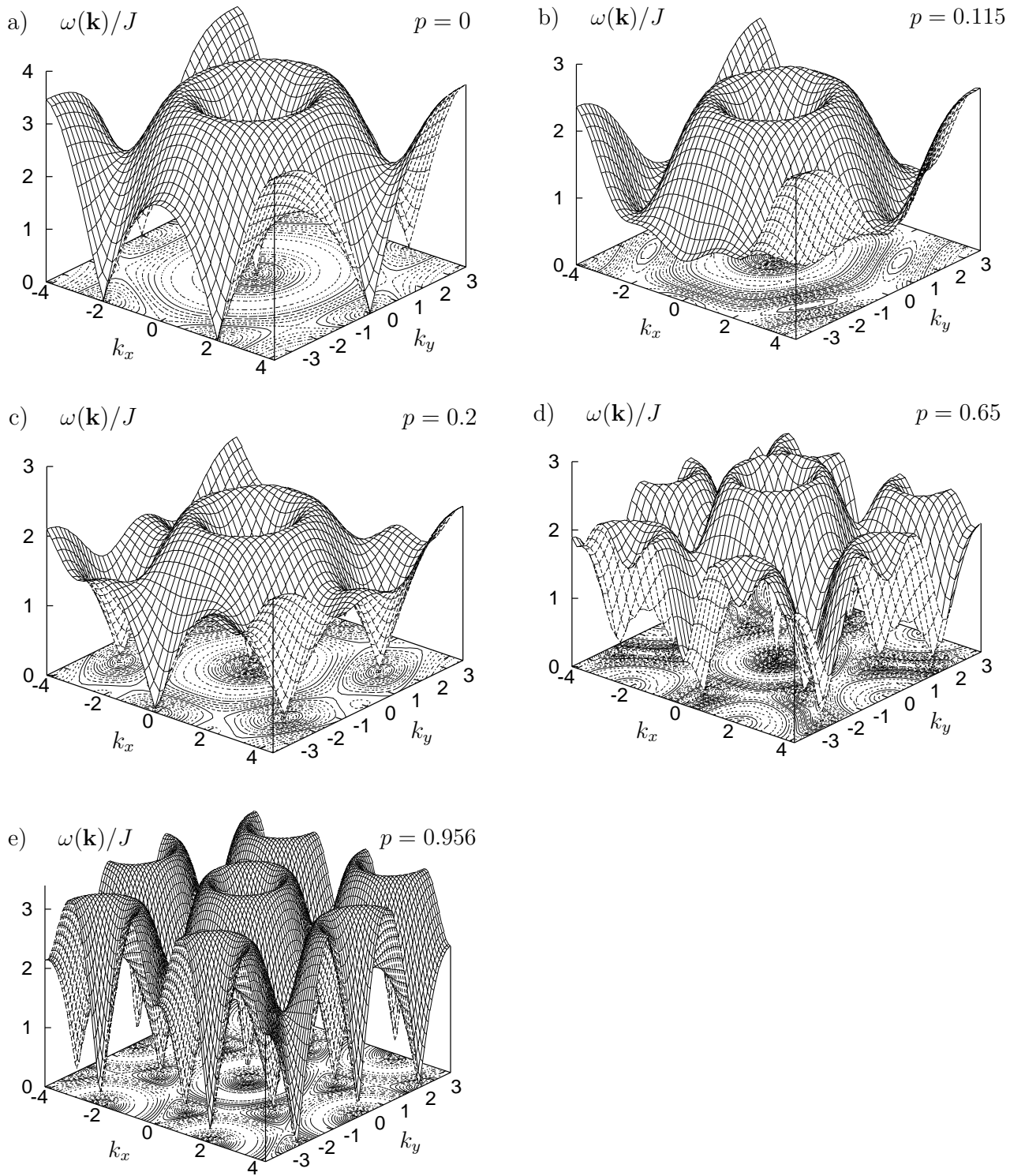}
\caption{ The dispersion of spin excitations $\omega_{\bf k}$ for different values of the frustration parameter $p$ at $T=0$. }
\label{fig2}
\end{center}
\end{figure}
\newpage
\begin{figure}[bhtp]
\begin{center}
\includegraphics[height=0.8\linewidth, angle=0]
{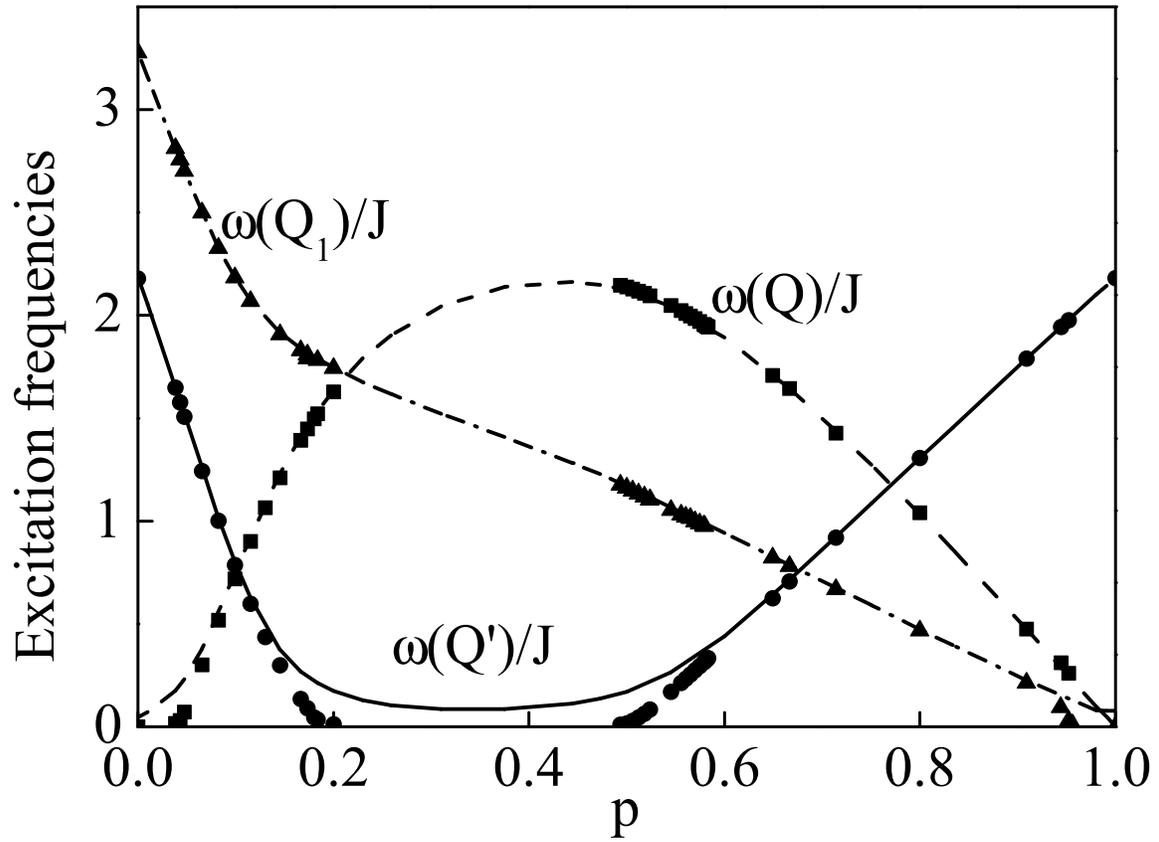}
\caption{The dependencies of the spin-excitation energies at the ordering wave vectors  ${\bf Q}$, ${\bf Q^\prime}$ and ${\bf Q}_1$ on the frustration parameter $p$ (symbols correspond to $T=0$ and lines to $T/J=0.2$). }
\label{fig3}
\end{center}
\end{figure}
\newpage
\begin{figure}[bhtp]
\begin{center}
\includegraphics[height=0.8\linewidth, angle=0]
{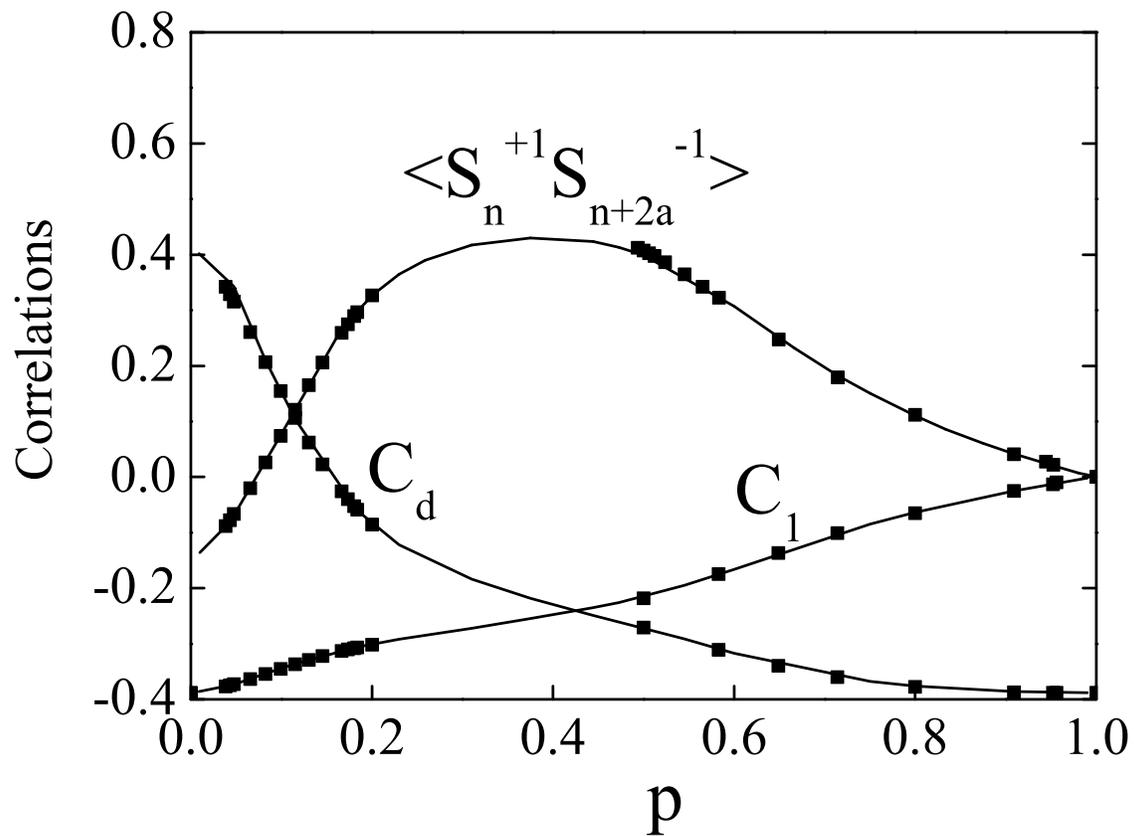}
\caption{The dependencies of the spin correlation functions between the nearest neighbors $(C_1)$,  the next-nearest neighbors $(C_d)$ and $\langle S^{+1}_{\bf n} S^{-1}_{\bf n+2a}\rangle$ on the frustration parameter $p$ at $T=0$ (squares) and $T/J=0.2$ (lines).}
\label{fig4}
\end{center}
\end{figure}
\newpage
\begin{figure}[bhtp]
\begin{center}
\includegraphics[height=0.8\linewidth, angle=0]
{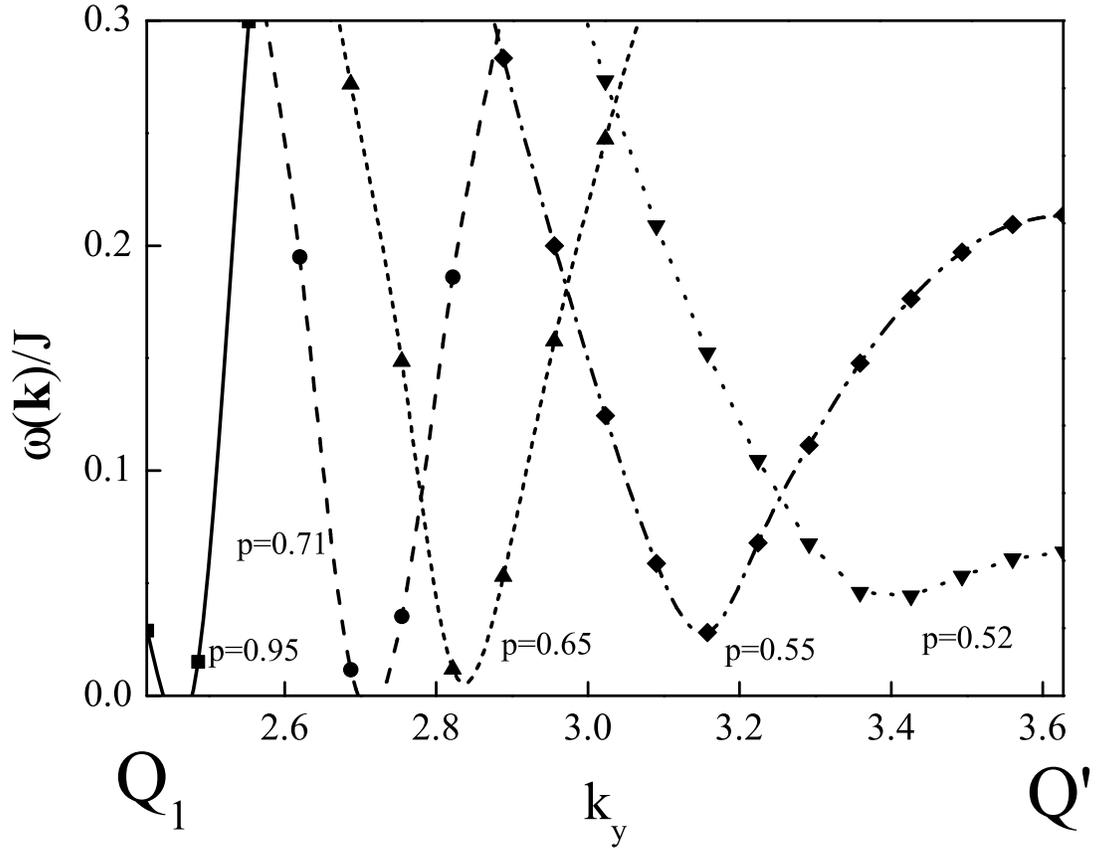}
\caption{ The dependencies of the spin-excitation frequency on the wave vector along the line ${\bf Q}_1-\bf Q^\prime$ for $p=0.52$, 0.55, 0.65, 0.71 and 0.95. $T=0$. The calculated values of the frequency are shown by symbols, lines are interpolations through these points.
}
\label{fig5}
\end{center}
\end{figure}
\end{document}